\documentclass[twocolumn,showpacs,amsmath,amssymb,pre]{revtex4}

\usepackage{graphicx}
\usepackage{dcolumn}
\usepackage{bm}

\begin{document}


\title{Ultracold neutral plasma expansion in two dimensions}

\author{E. A. Cummings}
 \altaffiliation[Present Address: ]{Lockheed Martin Space Systems Company, Sunnyvale,
 CA 94089}
\author{J. E. Daily, D. S. Durfee, and S. D. Bergeson}%
 \email{scott.bergeson@byu.edu}
 \affiliation{Brigham Young University, Department of Physics and
 Astronomy, Provo, UT 84602}

\date{\today}

\begin{abstract}
We extend an isothermal thermal model of ultracold neutral
plasma expansion to systems without spherical symmetry, and use this
model to interpret new fluorescence measurements on these
plasmas.  By assuming a self-similar expansion, it
is possible to solve the fluid equations analytically and
to include velocity effects to predict the fluorescence
signals.  In spite of the simplicity of this approach,
the model reproduces the major features of the experimental
data.
\end{abstract}

\maketitle

Ultracold plasmas are produced from photo-ionized laser-cooled
gases \cite{killian99,kulin00,killian01}.
In these laboratory plasmas, it is possible to study
the kinetics and thermodynamics of multi-component,
strongly-interacting Coulomb systems.  These systems are
characterized by the ratio of the nearest-neighbor
Coulomb energy to the average kinetic energy, denoted as
$\Gamma = (e^2/4 \pi \epsilon_0 d) / (k_b T)$, with $d$ being
the interparticle spacing.

A new class of ultracold plasma experiments has recently
become available in which it is possible to spectroscopically
study the plasma ions \cite{chen04,simien04,cummings05}.
These plasmas are made using alkaline-earth
atoms, because the resonance transition wavelengths of the
ions are readily
generated using standard laser methods.
The spatially-resolved time evolution of the plasma ion
temperature and density can be measured
using absorption and fluorescence techniques.

These ultracold neutral plasmas are not trapped, although efforts
are underway in a few laboratories to trap them.
The untrapped plasmas freely expand, and as they expand the density
and temperature change radically.  Processes of recombination,
collisional and thermal ionization, radiative cascade, and adiabatic and
evaporative cooling all play important roles in how the system
evolves and equilibrates.

A variety of models have been used to investigate the properties
of these plasmas
\cite{robicheaux02,robicheaux03,kuzmin02a,kuzmin02b,mazavet02,pohl04,pohl05}.
One particularly simple isothermal fluid model \cite{robicheaux02}
has been surprisingly successful in predicting the general features
of these plasmas \cite{robicheaux03,pohl04,cummings05}.
In this paper we extend this model from
the spherically-symmetric Gaussian plasma distributions to
Gaussian distributions with elliptical symmetry.

The elliptical symmetry has important experimental advantages.
In such systems the plasma expands primarily in two dimensions.  The
practical advantage is that the density falls more slowly than in the three
dimensional case, making it possible to study the plasma for longer
times.  The Doppler-shift due to the directed expansion
of the plasma is also suppressed.  It should therefore be possible
to study plasma oscillations and heating effects for greater time
periods before these
oscillations are masked by the directed expansion of the plasma.
Finally, if the plasmas are generated from a density-limited neutral
atom trap, the elongated symmetry allows a greater number of atoms to
be trapped initially, corresponding to a greater column density of
plasma ions.  This directly increases the visibility of fluorescence
and absorption signals.

\section{Isothermal fluid model}

An isothermal fluid model has been presented in the literature
\cite{robicheaux02,robicheaux03}.  It
successfully reproduces most of the major features of
recent experimental work.  This model was motivated by trends observed
in more sophisticated treatments.  The basic ideas of the model will
be reviewed here, and an extension to the case of a Gaussian
distribution with elliptical symmetry will be presented.

The initial ion density distribution is proportional to the
Gaussian distribution of the neutral atom cloud from which the
plasma is created.  For a spherically symmetric cloud, the
initial distribution can be written as $n(r)=n_0 \exp(-\beta
r^2)$. Because the electrons thermalize much faster than the ions,
in this model we take the initial electron density distribution
to be the thermal equilibrium distribution, given by the
Boltzmann factor:

\begin{equation}
n_e(r) = n_{0e} \exp\left[\frac{eV(r)}{k_B T}\right]
\end{equation}

\noindent The lowest temperature plasmas are nearly
charge-neutral, and the electron density distribution is
approximately equal to the ion density.  In this limit, it is
shown in \cite{robicheaux02,robicheaux03} that for a spherically
symmetric plasma that to within an arbitrary additive constant the
electrical potential energy can be written as

\begin{equation}
eV(r)  =  {k_B T} \ln[n(r)/n_{0e}]
 =  -     {k_B T \beta r^2} .
\end{equation}

\noindent The force is
the negative gradient of this potential energy.  It is manifestly
radial, spherically symmetric, and linearly proportional to
the radial coordinate $r$, measured from the center of
the plasma.  The velocity, which is the time integral
of the acceleration, is also linearly proportional
to $r$. The consequence is that
if the distribution is Gaussian initially, it will
remain Gaussian at all times in the expansion.

For the case of non-spherical symmetry, the approach is more or
less the same, although the isothermal nature of the
plasma has a more restricted meaning.  We will take the initial ion
density distribution to be Gaussian, symmetric in the $x-y$ plane, and
initially elongated in the $z$ direction:

\begin{equation}
n(r,t) = \frac{N \beta_1 \beta_2^{1/2}}{\pi^{3/2}}
\exp\left[- (x^2+y^2)\beta_1(t) - z^2\beta_2(t)\right]
\label{eqn:density} ,
\end{equation}

\noindent The initial conditions are $\beta_1(0)=\sigma_0^{-2}$ and
$\beta_2(0)=\alpha^2\sigma_0^{-2}$, and the parameter $\alpha$ defines
the elipticity of the system.  The plasma
fluid equations for our system
are written

\begin{eqnarray}
\frac{\partial n}{\partial t} + \nabla \cdot
\left(n \vec{v}\right) & = & 0 \label{eqn:continuity} \\
\frac{\partial \vec{v}}{\partial t} +
\left(\vec{v}\cdot\nabla\right) \vec{v} &=& \vec{a} \label{eqn:newton}.
\end{eqnarray}

Following the derivation used in the case of spherical symmetry,
the velocity is taken to be $\vec{v} = \gamma_1(t) (x \hat{x} + y \hat{y})
+ \gamma_2(t) z \hat{z}$.
Inserting this and the density profile of Eq. \ref{eqn:density} into
Eq. \ref{eqn:continuity} gives

\begin{eqnarray}
2\beta_2 \left( x^2\beta_1 +y^2 \beta_1 -1 \right)
\left(\dot{\beta}_1 + 2 \beta_1 \gamma_1 \right) \cdots & \nonumber \\
+ \;
4\beta_1\left( z^2 \beta_2 -1 \right)
\left(\dot{\beta}_2 + 2 \beta_2 \gamma_2 \right) & =0 .
\end{eqnarray}

\noindent Because $x$, $y$, and $z$ are independent variables, the only
non-trivial solution of this equation is

\begin{equation}
\gamma  =  - \dot{\beta} / 2\beta \label{eqn:gamma1},
\end{equation}

\noindent where we have dropped the subscripts because all
components have this same form of solution.

Solving Eq. \ref{eqn:newton} requires a little more care.  It is straight-
forward to write down the acceleration vector following the derivation
of the spherically symmetric case.  However, the solution requires that the
temperature be isothermal in a given dimension, but anisotropic in space.
This condition allows the density distribution to reduce to the proper form
in the limiting case of a plasma infinitely long in the $z$ dimension.  This
decouling requires energy to be conserved
separately in the $x-y$ plane and in the $z$ dimension. The
plasma equations for these two spaces are now exactly identical and
completely separable.

Equation \ref{eqn:newton} can be re-written (dropping the subscripts)
as

\begin{equation}
\dot{\gamma} + \gamma^2 = 2 k_b T(t) \beta(t)/m \label{eqn:velocity2},
\end{equation}

\noindent and the conservation of energy
is

\begin{equation}\label{eqn:energy}
    T(0) = T(t) + \frac{m}{2 k_b}\frac{\gamma^2}{\beta},
\end{equation}

\noindent where we have neglected the energy due to electron-ion
recombination.  Equations \ref{eqn:gamma1}, \ref{eqn:velocity2}, and
\ref{eqn:energy} are exactly identical to Eq. 2 of Ref.
\cite{robicheaux02}.  Using Eqs. \ref{eqn:gamma1} and \ref{eqn:energy},
we now have both $T$ and $\gamma$ in terms of $\beta$.  Inserting
this into Eq. \ref{eqn:velocity2} gives

\begin{equation}\label{eqn:beta}
    \frac{\ddot{\beta}}{\beta^2} - 2\frac{\dot{\beta}}{\beta^3} + 2v_e^2 = 0,
\end{equation}

\noindent where we have made the substitution $v_e^2=2k_b T(0)/m$.  The
solution to this equation is
$\beta_1^{-1} = \sigma_0^2 + ct + v^2 t^2$, where
$c$ is an integration constant.  The constant $c$ must be equal to
zero to meet the
condition that the ion velocity is initially zero at $t=0$
\cite{bergeson03}.  The time evolution of the density and velocity functions
can now be written as

\begin{eqnarray}
  n(\vec{r},t) & = & \frac{N
    \exp\left[- (x^2+y^2)/\sigma_1^2(t)
    - (z^2)/\sigma_2^2(t) \right]}
    {\pi^{3/2} \sigma_1^2(t) \sigma_2(t)} \label{eqn:dens} \\
  \vec{v}(\vec{r},t) & = & v_e^2 t
    \left( \frac{x}{\sigma_1^2(t)}\hat{x}
    + \frac{y}{\sigma_1^2(t)}\hat{y}
    + \frac{z}{\sigma_2^2(t)}\hat{z}
    \right) \label{eqn:bigVelocity}\\
  \sigma_1^2(t) & = & \sigma_0^2 + v_e^2t^2 \\
  \sigma_2^2(t) & = & \alpha^2\sigma_0^2 + v_e^2t^2 .
\end{eqnarray}

We note that for two-dimensional planar Gaussian charge
distributions, a closed-form expression for the electric field
has been derived \cite{bassetti80,furman94}. If such a compact
analytical solution could be written in the three-dimensional
case, it would remove the decoupling constraint that we imposed
in order to solve the plasma equations.  However, such a solution
is not readily apparent.

\section{Fluorescence signal model}

\begin{figure}
  \includegraphics[angle=270,width=3.4in]{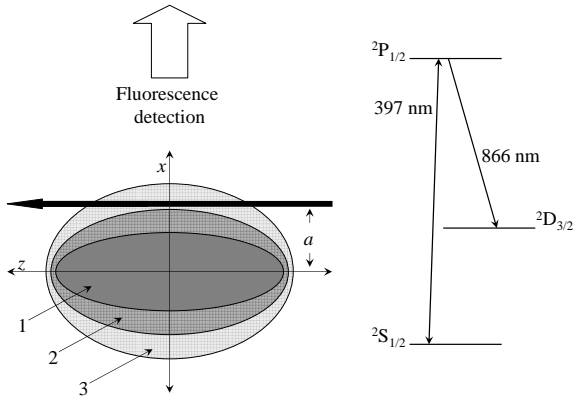}\\
  \caption{A schematic diagram of the fluorescence
  measurements.  {\it Left} --- The plasma is initially at a high
  density, and elongated in the $z$-dimension (labeled region 1).
  The probe laser is represented by the solid dark arrow parallel
  to the $z$ axis, displaced by a variable distance $a$.
  At later times the plasma expands slowly in $z$, but quickly
  in the $x-y$ plane.  Regions 2 and 3 represent the plasma size
  at later and later times.  As the plasma expands,
  the number of atoms in the column defined by the probe laser
  changes.  {\it Right} --- a partial level diagram of Ca$^+$.
  The probe laser is locked to the 397 nm resonance transition, and
  we measure
  397 nm light scattered by plasma ions.  }\label{fig:fluorescence}
\end{figure}

The geometry of our measurements is shown in Fig.
\ref{fig:fluorescence}.
In the experiment, the probe laser beam is spatially filtered and
focused into the plasma with a confocal beam parameter that is long
compared to all plasma dimensions.  The position offset of this
probe laser relative to the
plasma is denoted by the parameter $a$.  After the plasma
is created, the number of atoms in the column defined by the
probe laser beam changes dramatically.  When the laser beam  passes
through the center of the plasma ($a=0$), the number
of ions in the beam monotonically decreases.  However, if
the laser is outside of the initial plasma distribution, as
shown in Fig. \ref{fig:fluorescence}, the number first increases
as the plasma moves into the laser beam, and then decreases as
the plasma disperses.

The fluorescence signal depends on both the number of ions in
the column defined by the probe laser beam (Gaussian beam
profile with $1/e^2$ radius $w$) and the velocity
distribution of the ions in the plasma.  Because the laser
has a narrow bandwidth, atoms moving at velocities greater than
$\sim$9 m/s are Doppler-shifted out of resonance.  In this section
we will use the results of the previous section to derive
an expression for how the plasma ion fluorescence signal should
change with time for different values of the offset parameter $a$.

The fluorescence signal $s(t)$ is proportional to the absorption
of the probe laser beam.  Using Beer's law and the standard
approximation of small optical depth, $s(t)$
can be written as

\begin{equation}
 s(t) \propto  \int_{\mbox{\footnotesize{Vol}}}
 n(x,y,z,t)f_1(x-a,y)\bar{\sigma}(\nu-\nu_0), \label{eqn:sig}
\end{equation}

\noindent where $f_1$ is the spatial profile of the
probe laser beam, and $\bar{\sigma}$ is the absorption
cross section as a function of $\nu - \nu_0$, the difference
between the laser frequency and the atomic resonance frequency.
Removing the limitation of small optical depth is
trivial.
We can simplify this expression by setting the laser
frequency equal to $\nu_0$, and
recognizing $\nu - \nu_0 = v/\lambda$ as the Doppler shift
due to the velocity $v$ of the atoms, where $\lambda$ is
the optical wavelength of the
transition.   Equation \ref{eqn:bigVelocity} gives the
position-dependent velocity of the ions.  We define a length
$\ell \equiv {\Gamma \lambda \sigma_2^2}/{2 v_e^2 t}$, where
$\Gamma=1/2\pi \tau$ is natural width of the transition, $\tau$
is the lifetime of the transitions's upper state, and use
it to write the absorption profiles,

\begin{equation}
\bar{\sigma}/\bar{\sigma}_0 =
\left\{
  \begin{array}{ll}
   \left[\left(z/\ell\right)^2 +1\right]^{-1} & \mbox{Lorentzian, } \\ \\
   \exp\left[ -2 \left( {v_{th}}/{2 \Gamma \lambda}\right)^2 \left({z}/{\ell} \right)^2 \right] & \mbox{Gaussian,}
  \end{array}
\right.
\end{equation}

\noindent where $\bar{\sigma}_0$ is the absorption cross-section
on resonance and $v_{th}$ is the rms velocity of a
thermal distribution.  The true absorption lineshape is
better represented by a Voigt profile.  However, as the
Voigt profile can be approximated by a linear combination
of the Lorentzian and Gaussian line profiles \cite{liu01}, we will only
write down these two limiting forms.  Power broadening of
the line can be included in a straightforward manner \cite{citron77}.

We take the plasma density profile from Eq. \ref{eqn:dens}, and
write the spatial profile of the probe laser beam as

\begin{equation}
f_1 = \exp\left[ -2(x-a)^2/w^2 - 2y^2/w^2\right],
\end{equation}

\noindent which corresponds to the geometry represented in Fig.
\ref{fig:fluorescence}.  Performing the integration in Eq. \ref{eqn:sig}
gives,

\begin{equation}
 s(t) \propto \frac{\xi}{a^2}\exp(-\xi) \left\{
   \begin{array}{ll}
    {\eta} \; \mbox{erfc}(\eta) \; \exp(\eta^2) & \mbox{Lorentzian,} \\ \\
    1/\sqrt{ 1 + (2\Gamma \lambda / v_{th} \eta)^2 } & \mbox{Gaussian,}
   \end{array}
 \right.\label{eqn:final}
\end{equation}

\noindent where $\xi =  2a^2/(w^2 + 2\sigma_1^2)$ and
$\eta = \ell/\sigma_2$.  This expression for the Lorentzian
lineshape is proportional to Eq. 5 of Ref. \cite{cummings05}.

\section{Ultracold calcium plasmas}

\begin{figure}
\includegraphics[angle=270,width=3.4in]{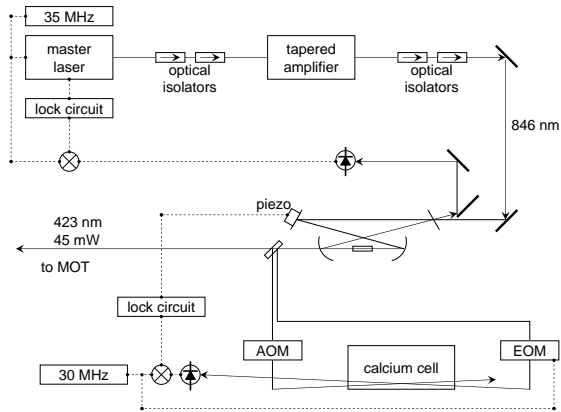}
\caption{\label{fig:exptSketch} A schematic drawing of the MOT
laser system and frequency stabilization electronics used in
these experiments.}
\end{figure}

We create ultracold neutral plasmas by photoionizing
laser-cooled calcium atoms in a magneto-optical
trap (MOT).  The calcium MOT
is formed in the usual way
by three pairs of counter-propagating
laser beams that intersect at right angles in the center of a
magnetic quadrupole field \cite{raab87}.  The 423 nm laser light
required for the calcium MOT is generated by frequency-doubling
an infrared laser in periodically-poled KTP (PPKTP),
and has been described previously
\cite{ludlow01}. A diode laser master-oscillator-power-amplifier
(MOPA) system delivers 300 mW single frequency at 846 nm, as
shown in Fig. \ref{fig:exptSketch}.  This laser is phase-locked to
a build-up cavity using the Pound-Drever-Hall technique
\cite{pound83}, giving a power enhancement of 30.  A 20mm long
PPKTP crystal in the small waist of the build-up cavity
is used to generate typically 45 mW output power at
423 nm \cite{gaudarzi03,targat04}.

The laser is further stabilized by locking the 423 nm light to the
calcium resonance transition using saturated absorption
spectroscopy in a calcium vapor cell \cite{libbrecht95}.  Our vapor
cell differs from Ref. \cite{libbrecht95} in that it
has a stainless steel body with conflat metal seals and
windows and a valve.  An
acousto-optic modulator (AOM) in one arm of the saturated
absorption laser beams shifts the laser frequency so that the
laser beam sent to the MOT is 35 MHz (one natural linewidth) below
the atomic resonance.  We also use the AOM to chop this beam and
use a lock-in amplifier to eliminate the Doppler background in the
saturated absorption signal.  Because the 846 nm laser is already
locked to the frequency-doubling cavity, the feedback from this
second lock circuit servos the frequency-doubling cavity length.

The trap is loaded from a thermal beam of calcium atoms that
passes through the center of the MOT.  The thermal beam is formed
by heating calcium in a stainless steel oven to
$650^{\mbox{\footnotesize{o}}}$ C. The beam is weakly collimated
by a 1mm diameter, 10mm long aperture in the oven wall.  The
distance between the oven and the MOT is approximately 10 cm.  As the
beam passes through the MOT, the slowest atoms in the velocity
distribution are cooled and trapped. An additional red-detuned
(140 MHz, or four times the natural linewidth) laser beam
counter-propagates the calcium atomic beam, significantly
enhancing the MOT's capture efficiency. To prevent optical pumping into
metastable dark states we also employ a diode laser at 672 nm.
The density profile of the MOT has an asymmetric Gaussian profile
and is well-represented by Eq. \ref{eqn:dens} with the peak
density equal to $4 \times 10^{9}$ cm$^{-3}$, $\sigma_0 = 0.5$ mm,
and $\alpha = 2.5$.

We photo-ionize the atoms in the MOT using a two-color, two-photon
ionization process.  A portion of the 846 nm diode laser radiation
from the MOT laser is pulse-amplified in a pair of YAG-pumped dye
cells and frequency doubled.  This produces a 3 ns-duration laser
pulse at 423 nm with a pulse energy around 1 $\mu$J.  This laser
pulse passes through the MOT, and its peak intensity is a few thousand
times greater than the saturation intensity.  A second YAG-pumped dye laser at
390 nm counter-propagates the 423 nm pulse and  excites the MOT
atoms to low-energy states in the region of the
ionization potential.  We
photo-ionize 85-90\% of the ground-state atoms in the MOT.  The
minimum initial electron temperature is limited by the
bandwidth of the 390 nm laser to about 1 K.

\begin{figure}
  \includegraphics[angle=270,width=3.4in]{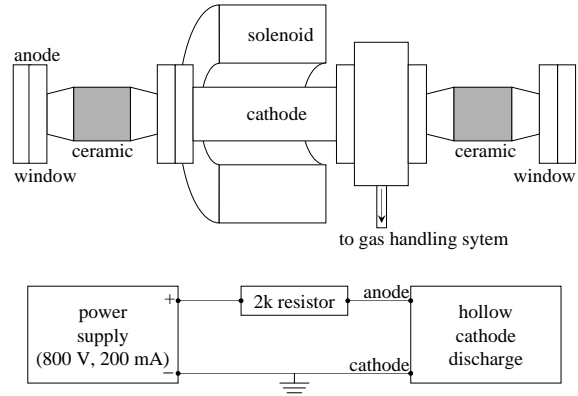}\\
  \caption{A schematic diagram of the DAVLL cell.
  {\it Upper} --- A drawing of the hollow.cathode discharge cell,
  approximately to scale.  The cathode is 10 cm long, with a
  1 cm diameter bore.  The operating pressure is between 100
  and 200 mTorr of krypton.
  {\it Lower} --- Electrical connections to the discharge cell.
  For increased current stability, a 2k$\Omega$ ballast resistor
  is connected in series with the discharge.
}\label{fig:cathode}
\end{figure}

Ions in the plasma scatter light from a probe laser beam tuned to
the Ca {\sc II} $^2S_{1/2} - ^2P_{1/2}$ transition at 397 nm.
The probe laser is generated by a grating-stabilized violet diode laser.
This laser, as well as the 672 nm laser used in the neutral atom
trap, is locked to the calcium ion
transition using the DAVLL technique \cite{corwin98}
in a large-bore, low-pressure hollow cathode discharge of our
own design (see Fig. \ref{fig:cathode}).

The probe laser is spatially filtered.  The typical probe laser
intensity is a few hundred $\mu$W
focused to a Gaussian waist of 130 $\mu$m in the
MOT.  We average repeated measurements of the scattered 397 nm
radiation with the probe laser in a given position, denoted
by the parameter $a$ in Fig. \ref{fig:fluorescence}.  This produces
a time-resolved signal proportional to the number of atoms
resonant with the probe beam in a particular column of the plasma.
By translating a mirror just outside the MOT chamber, we scan
the probe laser across the ion cloud.  In this manner we obtain
temporal and spatial information about the plasma expansion.

\section{Comparing the model to the data}

One comparison of the isothermal model with experimental data
was presented in Ref. \cite{cummings05}.  In that work,
the initial electron energy of the plasma, and therefore
the expansion velocity $v_e$, was fixed,
and the parameter $a$ was varied from 0 to 4$\sigma_0$.

\begin{figure}
 \includegraphics[width=3.4in]{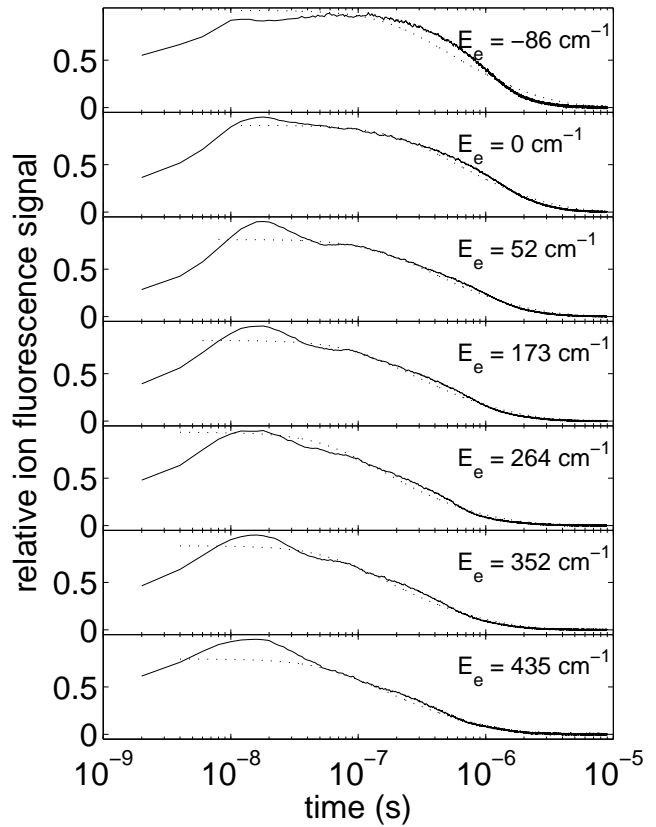}
 \caption{Relative ion fluorescence signal.  The probe laser
 is tuned to resonance and propagates through the center of the plasma
 ($a=0$). The electron energy, $E_e$, is measured relative to the
 ionization limit.  The solid line is the fluorescence signal,
 the dotted line
 is a fit using the Lorentzian lineshape in Eq. \ref{eqn:final}.
 The model is normalized to the fluorescence signal
 at $t=10^{-7}$ s, and the single
 fit parameter is the expansion velocity, $v_e$.
 In the top panel, the plasma is spontaneously
 generated after exciting the atoms to a Rydberg state with
 $n^* \sim 35.7$.} \label{fig:550}
\end{figure}

In the following, we present a complementary
comparison.  The solid line in Fig. \ref{fig:550} shows ion
fluorescence
signal with the probe laser tuned to the ion resonance frequency,
and with the probe laser beam propagating along the $z$ axis
($a=0$ in Fig. \ref{fig:fluorescence}) for a range of
initial electron energies.

As discussed in Ref. \cite{cummings05}, the early rise in the
fluorescence signal shows the increasing number of
plasma ions in the $^2P_{1/2}$ level.  This feature is easily
explained in terms of the classic Rabi two-level atom
with damping.  The probe laser intensity is in the
range of 5 to 10 times the resonance Rabi frequency.  Our
numerical integration of the
optical Bloch equations mimics the approximately 20 ns
rise time observed
in the experimental data, and shows that the
Gaussian spatial profile
of the probe laser beam washes out subsequent oscillations
in the excited state fraction.
Following the initial rise, the fluorescence decays.  At approximately
$t=50$ ns the decay slows down due to
correlation-induced heating in the plasma \cite{simien04,cummings05}.

\begin{figure}
 \includegraphics[width=3.28in]{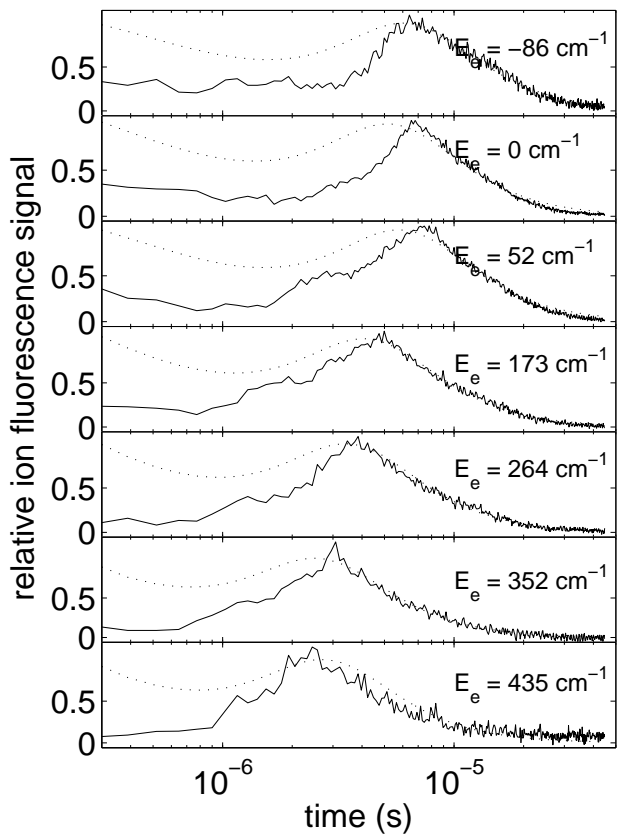}
 \caption{Relative ion fluorescence signal. The probe laser is
 tuned to resonance and propagates at a distance
 $a=1.1 \mbox{mm} = 2.2 \sigma_0$ relative to the center of the
 initial distribution.  The solid line is the data, and the
 dotted line is the fit using the Lorentzian lineshape in
 Eq. \ref{eqn:final}.  The model is normalized to the
 peak of the data.} \label{fig:600}
\end{figure}

\begin{figure}
 \includegraphics[width=3.28in]{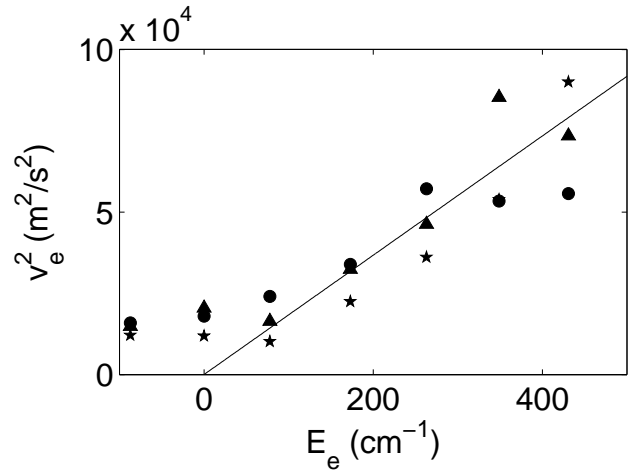}
 \caption{Expansion velocity extracted from Figs.
 \ref{fig:550} and \ref{fig:600}.  The triangles ($\blacktriangle$)
 represent the velocity extracted from Fig. \ref{fig:550},
 and the circles ({\large $\bullet$}) from Fig. \ref{fig:600}.  The
 stars ($\bigstar$) represent the velocity extracted from centering
 the peak of the model on the peak of the data in Fig. \ref{fig:600}. The line
 is a fit of the data with $E_e>150$ cm$^{-1}$.  The
 fitted result is $E_e = 3.0 m v_e^2$, where $m$ is the mass
 of the calcium ion.
 \label{fig:vel}}.
\end{figure}

These two processes are not included in the isothermal model.
We therefore begin the comparison of the model with the data
at time $t=10^{-7}$ s.  The model is normalized to the data
and fit using $v_e$ as the only fit parameter in an
un-weighted least-squares fitting routine.
The model uses the Lorentzian lineshape in Eq. \ref{eqn:final}.
The justification for using this lineshape arises from the
analysis presented in Ref. \cite{cummings05}.  In Fig. 2 of that
reference, the $t=0$ velocity is 6 m/s.   This
is the rms velocity of a Boltzmann distribution, and it is due
almost entirely to correlation-induced
heating.  This velocity width gives a Doppler width smaller than the
natural line width, and the Voigt profile is close to
a pure Lorentzian.  Furthermore, as the plasma evolves, the ion
temperature falls due to the adiabatic expansion.  The time scale
for this is $\sim \sigma_0 / v_e = 2 \;\mu$s.  This is shorter
than the time scale for heating the ions by collisions with the
electrons, $\sim m_{\mbox{\footnotesize{Ca}}}
/ m_e \omega_p = 50 \;\mu$s.  It is therefore not surprising
that the Gaussian lineshape gives a poorer fit to the data.  In
the fitting procedure, the Gaussian lineshape
produces expansion velocities that do not grow as the initial
electron energy increases.
The figure shows that the Lorentzian model describes
the data well over a few orders of magnitude in time.  The velocities
extracted from these data are shown in Fig. \ref{fig:vel}.

The differences between the model and the signal are not
negligible.  For all initial
electron energies, the model is slightly too low at
$t=300$ ns, and too high at $t>1 \;\mu$s.  The data
are not corrected for optical pumping into the $D-$states,
which we measure to have a time constant of about 5 $\mu$s.
The differences between the model and the fluorescence
signal could indicate internal heating processes that
manifest themselves in
the ion velocity on the few hundred ns time scale.
They could also indicate ions that appear in the plasma from
Rydberg states at late times or collective plasma density
variations.  These differences can be studied in future work.

We also compare the fluorescence signal and the model
over a range of initial electron energies with the
probe laser beam shifted to  $a=1.1 \mbox{mm} = 2.2 \sigma_0$.
In this arrangement, the ion fluorescence signal is initially
small, and grows as ions move into the probe laser beam.
Typical data are shown in Fig. \ref{fig:600}.  For
these data, the model is fit to the $t>6\;\mu$s signal
using a least-squares procedure, with $v_e$ as the fit parameter.
The velocity extracted from this fit is shown in
Fig. \ref{fig:vel}.
It is possible to fit the data using $v_e$ so that the
peak in the model coincides with the peak of the
fluorescence signal.  These velocities are also plotted
in the figure.

For these data, the signal in the first $\mu$s is small
compared to the model.
Moreover, there are variations in the fluorescence signal
that do not appear in the model.
This suggests that at the edges of the plasma expansion,
the density distribution is distinctly non-Gaussian,
even at early times before any ion motion is possible.
It also appears, as pointed out in Ref. \cite{cummings05},
that the Gaussian density profile is recovered
at late times for all initial electron energies.

\section{conclusion}

We present an extension of the isothermal plasma expansion
model of Refs. \cite{robicheaux02,robicheaux03} for quasi-two-dimensional
geometry.  We include velocity effects and predict a fluorescence
or absorption signal vs. time for given initial conditions.
The model matches the correct order of magnitude and general
features of the fluorescence signal.  Some discrepancies are
pointed out, which can be studied in future work.

Making the plasmas more ideally two-dimensional should improve
applicability of the model, and further suppress effects due to
expansion in the long dimension.  Increasing the range of initial
electron temperatures and plasma densities can probe
interesting regions of phase space where differences between
the data and the model are likely to be more pronounced.

As an example, it should be possible to extend this model
to include effects
due to electron-ion recombination at early times.  For strongly-coupled
neutral plasmas, the three-body recombination rate should be on the
order of the plasma frequency.  Using the high-sensitivity and fast
time-response of fluorescence spectroscopy, it should be possible
to directly measure spectroscopic recombination signatures in
low-temperature, low-density plamsas, where the predicted
three-body recombination
rate is greater than the plasma frequency.  For example,
reducing the plasma density to 10$^6$ cm$^{-3}$
will make the recombination time $ \sim 1/\omega_p = 20$ ns.
Increasing the probe laser intensity
to 100 times the saturation
intensity will shorten the early rise time of the fluorescence
signal to around 10 ns.  Optical
pumping time will be comparable to the correlation-induced
heating time, around 200 ns.

\section{Acknowledgements}

This research is supported in part by Brigham Young University, the
Research Corporation, and the National Science Foundation (Grant No.
PHY-9985027).  One of us (SDB) also acknowledges the support
of the Alexander von Humboldt foundation.

\end{document}